\documentclass[onecolumn,secnumarabic,amssymb, nobibnotes, aps, prd,preprint,showpacs,]{revtex4-1}

\usepackage{epsfig}
\begin{document}

\markboth{}
{\textit{D.A. Fagundes, M.J. Menon}}

\title{Total Hadronic Cross Section and the Elastic Slope: An Almost Model-Independent Connection}

\author{D.A. Fagundes\footnote{fagundes@ifi.unicamp.br}, M.J. Menon\footnote{menon@ifi.unicamp.br}}

\affiliation{Universidade Estadual de Campinas - UNICAMP\\
Instituto de F\'{\i}sica Gleb Wataghin, 
13083-859 Campinas, SP, Brazil\\}
\pacs{13.85.-t, 13.85.Tp\\\\\\
\centerline{\textit{To be published in Nuclear Physics A} }
\vspace{0.5cm}
\centerline{(doi: 10.1016/j.nuclphysa.2012.01.017)}}

\begin{abstract}
An almost model-independent parametrization for the ratio of the total cross
section to the elastic slope, as function of the center of mass energy, is
introduced. The analytical result is based on the approximate relation of this
quantity with the ratio $R$ of the elastic to total cross section
and empirical fits to the $R$ data from proton-proton scattering above 10 GeV,
under the conditions of asymptotic unitarity and the black-disk limit.
This parametrization may be useful in studies of extensive air showers
and the determination of the proton-proton total cross section from 
proton-air production cross section in cosmic-ray experiments.
\end{abstract}

\maketitle

\noindent

\section{Introduction}

In addition to their intrinsic astrophysical importance, cosmic-ray experiments
constitute a valuable tool for the investigation of particle
and nuclear physics at energies far beyond those obtained
in accelerator machines. However, at the highest energies
a direct approach to particles
properties and  their interactions is difficult
due to the decreasing flux with the increase
of the energy. Presently, an indirect method,  based on extensive air shower 
(EAS) studies, is the usual way to treat the subject \cite{ralf}.
In these events,  the distribution of
the first interaction point allows, in principle, the determination
of the proton-air production cross section \cite{sokol} and in a second step, the
estimation of the most fundamental quantity in hadronic interactions: the proton-proton
total cross section \cite{egls,engel}.

However, in practice, the interpretation of the EAS development depends on extrapolations
from phenomenological models that have been tested only in the accelerator energy region, resulting 
therefore in systematic theoretical 
uncertainties.
That represents a crucial point because different models
with distinct physical pictures and at the same time consistent with the experimental data up to c.m. energies $\sim$
2 TeV, present, in general, contrasting extrapolations at the cosmic-ray region (above $\sim$ 50 TeV).
Therefore the theoretical uncertainties (bands) involved are relatively large and very difficult to be estimated. 
As a particular consequence, the estimations of the proton-proton total cross section
from different cosmic-ray experiments and analyses are characterized by large
error bars and even discrepant results, as discussed in \cite{alm} and references
therein.

EAS studies are essentially based on Glauber's multiple diffraction theory \cite{glauber,gm}
and its extensions and/or corrections, including Gribov-Regge screening effects and other ingredients.
In this context the evaluation of the hadron-nucleus elastic and quasi-elastic cross
sections, which contribute to the nucleon-air production cross section, depends
on the ratio of two physical quantities, the total cross section 
$\sigma_{tot}$ and the elastic slope $B$. That ratio
just represents one of the main sources of 
uncertainties in model extrapolations.

In this work an almost model-independent parametrization for
the ratio $\sigma_{tot}$/$B$ is proposed, which may avoid uncertainties from models
tested only at lower energies.
The parametrization is based on the approximate, but experimentally
justified, connection of  $\sigma_{tot}$/$B$ with the
ratio $R$ of elastic to total cross section. 
By means of
unitarity arguments
and empirical fits to $R$ data from $pp$ scattering above 10 GeV and up to
7 TeV, an analytical parametrization for $R(s)$ is introduced and then
extended to the ratio $\sigma_{tot}$/$B$.

The manuscript is organized as follows. In Sect. II we introduce the physical quantities
of interest with explicit reference to the importance of the ratio 
$\sigma_{tot}$/$B$ in cosmic-ray studies. In Sect. III we recall some
formal (rigorous) results from axiomatic QFT and how some inequalities can
be connected with experimental results. In Sect. IV we introduce the analytical parametrization
and present the fit results. The conclusions and some final remarks are the contents of Sect. V.

\section{Physical Quantities and the Glauber Formalism}

Let us first recall the main physical quantities related to
high-energy elastic hadron scattering, defining the notation 
and normalizations \cite{matthiae}. Neglecting spin effects 
and denoting $F(s,t)$ the invariant elastic amplitude in terms of the
Mandelstam variables, $s$ and $t$,
the differential and total cross sections at high energies ($s >>$ 1 GeV$^2$)
are expressed, respectively, by

\begin{eqnarray}
\frac{d\sigma}{dt}(s,t) = \frac{16\pi}{s^2}\, |F(s,t)|^2, 
\end{eqnarray}
and
\begin{eqnarray}
\sigma_{tot}(s) = \frac{16\pi}{s}\, {\rm Im}\, F(s, t=0)
\qquad
(\mathrm{Optical\ Theorem}).
\end{eqnarray}
The parameter $\rho$, the ratio between the real and imaginary parts of the
forward amplitude, is given by
\begin{eqnarray}
\rho(s) = \frac{{\rm Re}\, F(s,\ t=0)}{{\rm Im}\, F(s,\ t=0)},
\end{eqnarray}
and the slope of the elastic differential cross section in the forward direction
is defined as
\begin{eqnarray}
B(s,t=0) = \left[ \frac{d}{dt} \left(\ln \frac{d\sigma}{dt} \right) \right]_{t=0}.
\end{eqnarray}
From (1) to (3), the optical point is expressed by
\begin{eqnarray}
\left.\frac{d\sigma}{dt} \right|_{t=0} = 
\frac{\sigma_{tot}^2 (1 + \rho^2)}{16\pi}. 
\end{eqnarray}

The integrated elastic cross section reads
\begin{eqnarray}
\sigma_{el}(s) = \int_{t_0}^{0}\,\frac{d\sigma}{dt}(s,t) dt,
\end{eqnarray}
where $t_0$ defines the physical (kinematic) region
and, from unitarity, the inelastic cross section is obtained:
\begin{eqnarray}
\sigma_{in}(s) = \sigma_{tot}(s) - \sigma_{el}(s).
\end{eqnarray}

For our purposes let us recall two formulas in the Glauber formalism that play a central
role in EAS studies \cite{ralf}. The first one is the expression for the sum of the
elastic and quasi-elastic cross section for hadron-nucleus (hA) scattering,

\begin{eqnarray}
\sigma_{el}^{hA}(s) + \sigma_{qel}^{hA}(s) =
\int d^2b \left| 1 - \prod_{j=1}^{A} [1 - a_j(s,\vec{b} - \vec{b}_j)] \right|^2\,
\left[\prod_{j=1}^{A} \tau(\vec{r}_j)d^3r_j \right],
\end{eqnarray}
where $\vec{r}_j$ and $\vec{b}_j$ are the coordinate and impact parameter of the individual nucleons, $\tau(\vec{r}_j)$
the single nucleon density, $\vec{b}$ the impact parameter of the cosmic-ray hadron
and $a_j(s, \vec{b} - \vec{b}_j)$ the nucleon-nucleon impact parameter amplitude
(profile function). In addition to possible configurations for the nucleus, the 
profile function constitutes the main ingredient for the connection between hadron-hadron and
hadron-nucleus scattering. Typically this profile is parametrized by \cite{ralf}

\begin{eqnarray}
a_j(s,\vec{b}_j) = \frac{[1 + \rho(s)]}{4\pi}\, \frac{\sigma_{tot}(s)}{B(s)}\, 
e^{-\vec{b}_j^{2}/[2B(s)]},
\end{eqnarray}
where $\rho$, $\sigma_{tot}$ and $B$ are the quantities defined above,
demanding inputs from models to complete the connection. As clearly illustrated
by Ulrich \textit{et al}. \cite{ralf}, the uncertainty bands for these three quantities
resulting from high-energy extrapolations
based on representative phenomenological models, are larger than the range covered
by all available MC interaction model results, as QGSJET01c, EPOS1.61, SIBYLL2.1 and QGSJETII.3,
leading the authors to the conclusion that presently,  ``the extrapolation of hadronic cross sections to cosmic-ray energies might be underestimated'' \cite{ralf}.

In what concerns the above three fundamental quantities, we recall that forward amplitude analyses
connect $\sigma_{tot}(s)$ and $\rho(s)$ through dispersion relations. Detailed
tests on different parametrizations have been developed by the COMPETE Collaboration,
with the selection of the highest rank result \cite{compete1,compete2}, which also
appears in the Review of Particle Physics by the Particle Data Group \cite{pdg}.
Recent results by the TOTEM Collaboration  
on $\sigma_{tot}$ at 7 TeV \cite{totem} and the expected estimation of $\rho$ 
at this energy, will certainly shed light on novel 
analytical parametrizations and therefore more reliable extrapolations
with less model dependency.

However, that is not the case for the elastic slope $B(s)$ since the available data
from $pp$ and $\bar{p}p$ elastic scattering can be extrapolated in a very large
band of possibilities and moreover, any result is strongly model dependent. As a consequence,
from Eq. (9), despite the uncertainty in the effective radius of the nucleon-nucleon amplitude, 
any extrapolation is strongly dependent on the ratio
\begin{eqnarray}
\frac{\sigma_{tot}}{B}(s),
\end{eqnarray}
namely the unknown correlation between $\sigma_{tot}$ and $B$ in terms of energy.
Although some analytical connections have already been investigated 
from fits to the experimental data \cite{mmm},
the statistical and systematic errors in both quantities and the model
dependencies involved put limits on these results.

Based on the above comments, we understand that even under some reasonable approximate
conditions, an almost model-independent parametrization for the above ratio
may reduce the uncertainty band in the extrapolations from accelerator to 
cosmic-ray-energy regions. That is the point we are interested in here.

\section{Formal and Experimental Results}

In this section we first recall some rigorous (formal) results related
to $\sigma_{tot}(s)$ and $B(s)$ and their connections with the experimental data
presently available. Based on these considerations in the next section we introduce
our proposed parametrization and present the results.

\subsection{Rigorous Results}

General principles and high-energy theorems 
have always been a fruitful  source of model-independent results
for physical quantities in the asymptotic regime \cite{eden}.
In this context, two well-know inequalities have been derived for the total cross section
and the elastic slope. The first one is the Froissart-Lukaszuk-Martin
upper bound, stating that asymptotically ($s \rightarrow \infty$) 

\begin{eqnarray}
\sigma_\mathrm{tot}(s) \leq \frac{\pi}{m_{\pi}^2}\, \ln^2 \frac{s}{s_0},
\end{eqnarray}
for some $s_0$ \cite{fro,mar,luk}. The second one, playing here an important role, is the lower bound of MacDowell and Martin, 
obtained from unitarity together with properties of the Legendre polynomials and involving
forward quantities \cite{mac},

\begin{eqnarray}
2\left[\frac{d}{dt} \ln \mathrm{Im} F(s,t) \right]_{t=0}
\geq \frac{1}{18\pi} \frac{\sigma_{tot}^2(s)}{\sigma_{el}(s)}.
\end{eqnarray}
From the definition of the forward slope, Eq. (4), together with Eq. (1)
and under the assumption
\begin{eqnarray}
\mathrm{Re}\, F(s, t=0) < < \mathrm{Im}\, F(s, t=0),
\nonumber
\end{eqnarray}
it follows that

\begin{eqnarray}
\left.\frac{d}{dt} \ln \mathrm{Im} F(s,t)\right|_{t=0} \approx \frac{1}{2} B
\nonumber
\end{eqnarray}
and from (12) an upper bound is obtained for our ratio of interest,

\begin{eqnarray}
\frac{\sigma_{tot}(s)}{B(s)} \leq 18 \pi \frac{\sigma_{el}(s)}{\sigma_{tot}(s)}.
\end{eqnarray}

Although rigorous, relations involving inequalities have a limited practical applicability,
except for bounds imposed on the construction of phenomenological models. To go further in the 
search for almost empirical or model-independent results, experimental data and formal inequalities 
must be checked, as follows.

\subsection{Experimental Results}

The highest energies reached in accelerator machines for particle and antiparticles reactions concern
$pp$ and $\bar{p}p$ scattering, covering the region up to
$\sim$ 2 TeV ($\bar{p}p$) and, presently, up to 7 TeV ($pp$). These data indicate that at the highest energies
\begin{eqnarray}
\rho(s) \lesssim \mathrm{0.14},
\nonumber
\end{eqnarray}
which means that the above assumption, $\rho < < $ 1, constitutes a considerable approximation.
On the other hand, at the optical point, Eq. (5), an assumption like
\begin{eqnarray}
1 + \rho^2 \approx \mathrm{1}
\end{eqnarray}
certainly represents a reasonable approximation. We shall return to this point in what follows.
It may be interesting to  note that these information allow us to derive bound (13) from (12) 
under different assumptions, as shown in Appendix A.

Concerning the differential cross section, experimental data indicate a sharp forward peak,
followed by a dip-bump or dip-shoulder structure above $\sim$ 0.5 GeV$^2$ (Tevatron, LHC).
Typically, these structures are located more than 5 decades below the optical point, Eq. (5).
These experimental facts are important in the determination of the integrated elastic
cross section, since in this case the differential cross section can effectively be
represented by an exponential fall off, simulated by a model-independent parametrization \cite{totem},

\begin{eqnarray}
\frac{d\sigma}{dt} = \left.\frac{d\sigma}{dq^2} \right|_{q^2=0}\, e^{Bt},
\end{eqnarray}
with $B$ the (constant) forward slope. In that case, with the reasonable approximation (14)
at the optical point (5) and assuming $t_0 \rightarrow - \infty$ in Eq. (6),
the integrated elastic cross section reads

\begin{eqnarray}
\sigma_{el}(s) = \frac{1}{B(s)}\, \frac{\sigma_{tot}^2(s)}{16 \pi}
\nonumber
\end{eqnarray}
and therefore,
\begin{eqnarray}
\frac{\sigma_{tot}(s)}{B(s)} = 16 \pi \frac{\sigma_{el}(s)}{\sigma_{tot}(s)},
\end{eqnarray}
which is very close to the approximate bound (13).
However, the main ingredient in this result is the possibility to
investigate the behavior of $\sigma_{tot}(s)/B(s)$ from formal and experimental
information on the ratio $\sigma_{el}(s)/\sigma_{tot}(s)$, as discussed
in what follows.

\section{Analytical Parametrization and Fit Results}

In Figure 1 we display the experimental information presently available on the ratio
$\sigma_{el}/\sigma_{tot}$ from $pp$ scattering above 10 GeV \cite{pdg}, including the recent
TOTEM result at 7 TeV \cite{totem} (highest energy reached in accelerators). From
a strictly empirical point of view, the data in the linear-log scale may suggest a parabolic
parametrization in terms of $\ln s$, with \textit{positive} curvature. 
However, unitarity demands an \textit{obvious bound},
\begin{eqnarray}
\frac{\sigma_{el}}{\sigma_{tot}} \leq 1.
\nonumber
\end{eqnarray}
In addition, naive models, as for a Gaussian profile or the gray-disk, predict \cite{block}
$\sigma_{el}/\sigma_{tot} = C/2$, where $C$ is a constant (absorption coefficient) and
in the black-disk limit $C = 1$. These results indicate a constant asymptotic 
limit for the ratio
\begin{eqnarray}
\lim_{s \rightarrow \infty}\, \frac{\sigma_{el}}{\sigma_{tot}} = A \quad (\mathrm{constant})
\nonumber
\end{eqnarray}
and therefore a \textit{change of sign} in the curvature, at some finite
value of the energy, is expected.
Moreover, since from Fig. 1 the data at low energies indicate
$\sigma_{el}/\sigma_{tot}$ $\sim$ constant $\approx$ 0.18, 
a general behavior related to a logistic or sigmoid function
can be conjectured, at least above 10 GeV (the high-energy region).
Several functions with this property can be considered. However, for tests on goodness of fit some quantitative
information on the value of the asymptotic limit $A$ is necessary.

Looking for a wide range of possibilities in the phenomenological context, we shall consider
two contrasting pictures that have been discussed in the literature. On the one hand, the amplitude
analysis by Block and Halzen favour the asymptotic black-disk \cite{bh}, which is also
predicted, for example, in the models by Chou and Yang \cite{cy}
and by Bourrely, Soffer and Wu \cite{bsw}. On the other hand, the 
U-matrix unitarization scheme by Troshin and Tyurin predicts
$\sigma_{tot}(s) \sim \sigma_{el}(s) \sim \ln^2{s}$ and $\sigma_{inel}(s) \sim \ln{s}$ \cite{tt},
which is beyond the black-disk limit and in agreement with the above obvious unitarity bound.
Therefore these two contrasting pictures suggest
\begin{eqnarray}
A = \frac{1}{2} \quad (\text{black-disk\ limit})
\quad \mathrm{and} \quad
\frac{1}{2} < A \leq 1 \quad (\text{beyond\ the\ black-disk\ limit}).
\nonumber
\end{eqnarray}
Although, in principle, it might be possible to explore all the real interval
for $A$ beyond the black-disk, we consider here only its maximum value. As we shall show,
that is adequate and sufficient for our purpose to infer wider uncertainty bands from 
the above mentioned phenomenological context.

Based on these considerations and inspired
in different physical phenomena,
we have tested several functional forms to fit the  
 ${\sigma_{el}}/{\sigma_{tot}}$ data.
The best statistical result has been obtained with
the following novel model-independent parametrization:

\begin{eqnarray}
\frac{\sigma_{el}}{\sigma_{tot}}(s) = A \tanh (\gamma_{1} + \gamma_{2} \ln s + \gamma_{3} \ln^{2} s ),
\end{eqnarray}
where $\gamma_i$, $i = $ 1, 2, 3 are free fit parameters and $A$ represents the
asymptotic limit, for which we consider only the two extreme cases
$A = 1/2$ and $A = 1$.

The data reductions have been performed with the objects of the class TMinuit of ROOT Framework \cite{root}. We have adopted 
a Confidence Level  of $\approx$ 68 $\%$ (one standard deviation),
which means that the projection of the $\chi^{2}$ distribution in $(N+1)$-dimensional space ($N =$ number of free fit parameters) contains 68 $\%$ of probability \cite{bev}.
The fit results for $A = 1/2$ and $A = 1$ are displayed in Fig. 2 and Table I together with the statistical information.
In both cases, the error propagation from the fit parameters has been
evaluated and are also displayed in the figure; the bands however are  indistinguishable.

\begin{figure}[pb]
\centering
\epsfig{file=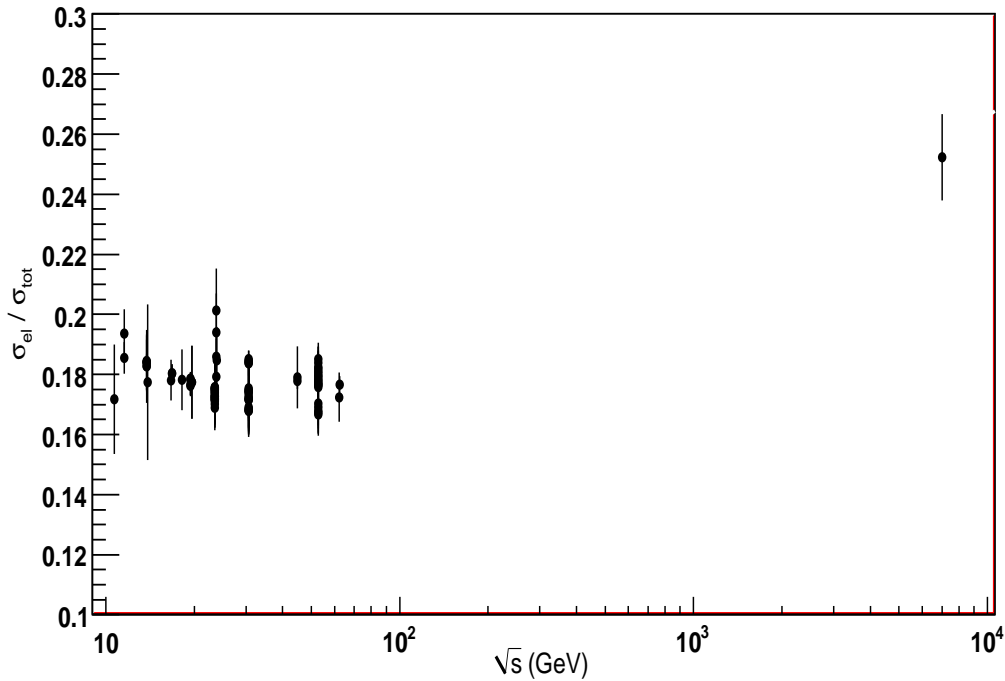,width=15cm,height=10cm}
\caption{Experimental data on the ratio between the elastic and total
cross sections
from $pp$ scattering above 10 GeV \cite{pdg,totem}.}
\label{f1}
\end{figure}
\begin{figure}[pb]
\centering
\epsfig{file=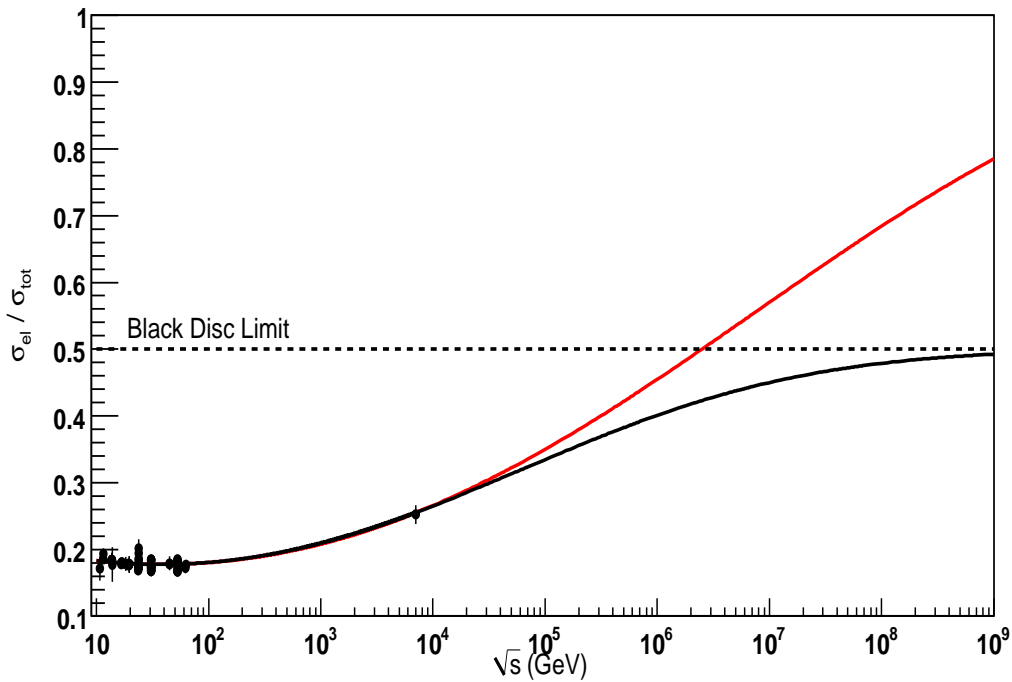,width=15cm,height=10cm}
\caption{Ratio between the elastic and total cross section and fit results through
parametrization (17), including uncertainty from error propagation, 
for $A=1$ (upper curve) and $A=1/2$ (lower curve).}
\label{f2}
\end{figure}

From the approximate result (16), the ratio $\sigma_{tot}/B$
can be predicted as function of the energy and in an almost model-independent
way. The results together with the experimental data \cite{pdg,totem,durham} are displayed in Fig. 3
and show that, in fact, Eq. (16) is very close to the approximate bound (13).
Up to our knowledge, the only rigorous result indicating a constant asymptotic value
for this ratio appears in the recent formal analysis by Azimov, on boundary values for the
physical cross section and slope \cite{azimov}. The numerical predictions with uncertainties for the ratio 
$\sigma_{tot}/B$ at the LHC energy region are displayed in 
Table II, together with the experimental value at 7 TeV. 

\begin{table}[ht]
\caption{Fit results with parametrization (17) for the ratio $\sigma_{el}/\sigma_{tot}$
from $pp$ scattering above 10 GeV. In both cases the degrees of freedom
($DOF$) are 87.}
\begin{center}
\begin{tabular}{ccc}
\hline
& A = 1/2 & A = 1 \\
\hline
$\gamma_{1} (\times 10^{-1})$&  4.66$\pm$ 0.18 &  2.204$\pm$ 0.078\\
$\gamma_{2} (\times 10^{-2})$& -2.59$\pm$0.49  & -1.11$\pm$0.20\\
$\gamma_{3} (\times 10^{-3})$&  1.77$\pm$0.33 &  0.76 $\pm$0.13 \\
\hline
$\chi^{2}/DOF$ & 1.167& 1.168\\
\hline
\end{tabular} 
\end{center}
\end{table}

\begin{figure}[pb]
\centering
\epsfig{file=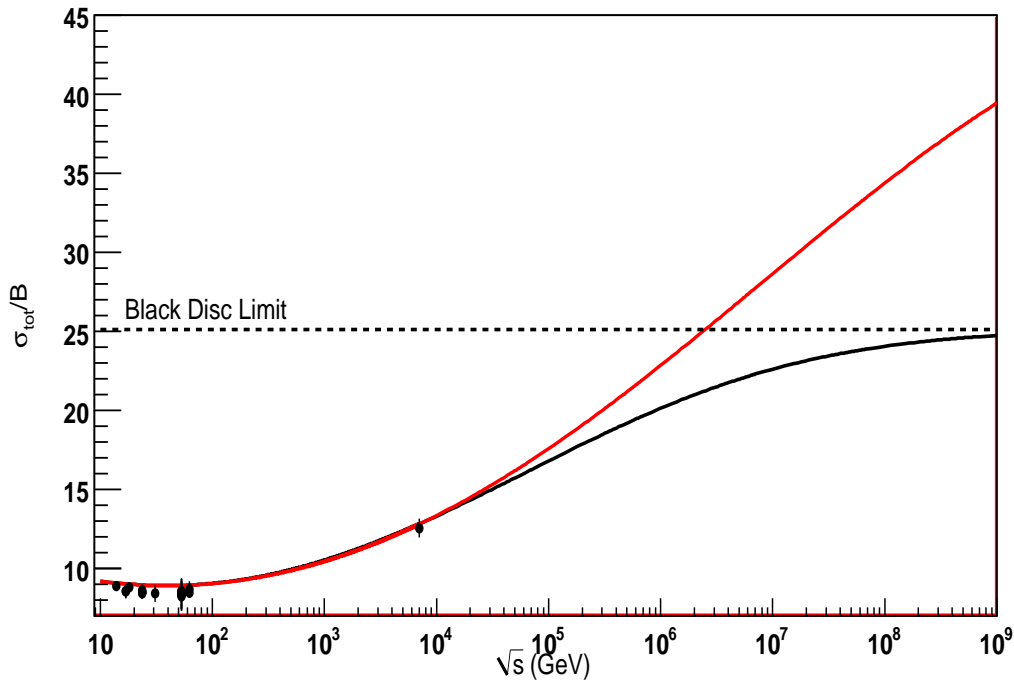,width=15cm,height=10cm}
\caption{ Experimental data on the ratio between the total cross section 
and the elastic slope \cite{pdg,totem,durham} and
predictions from Eqs. (16-17), including uncertainty from error propagation, 
for $A=1$ (upper curve) and $A=1/2$ (lower curve).}
\label{f3}
\end{figure}
\begin{figure}[pb]
\centering
\epsfig{file=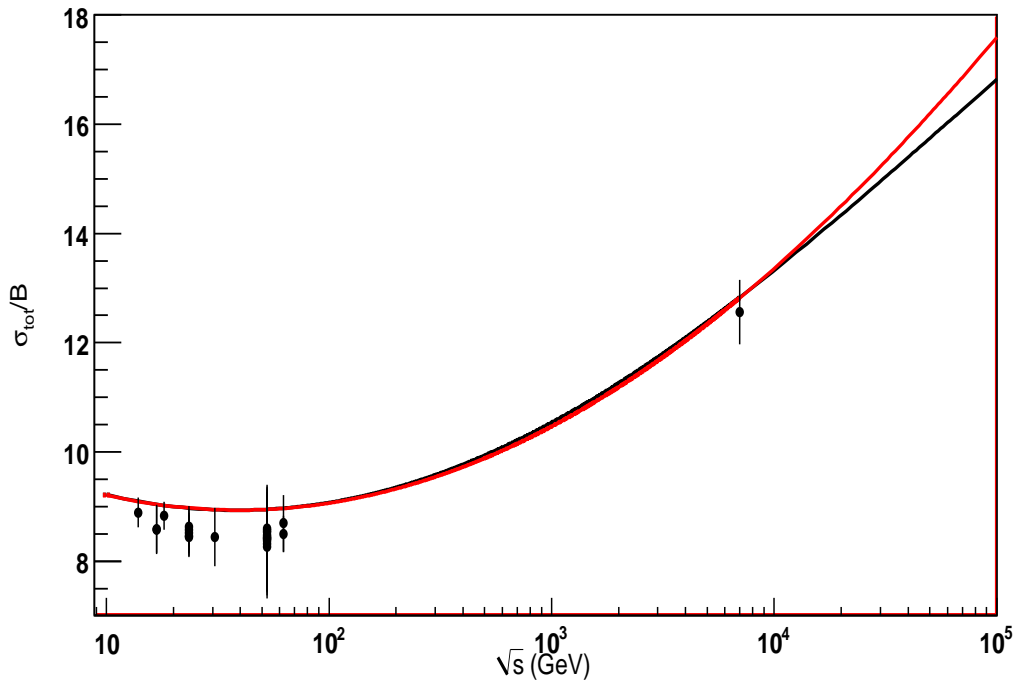,width=15cm,height=10cm}
\caption{Detail of Fig. 3 up to the Auger energy region.}.
\label{f4}
\end{figure}

\begin{table}[h]
\caption{Predictions from Eqs. (16-17) for the ratio $\sigma_{tot}/B$ at the LHC energy 
region and the TOTEM result at 7 TeV \cite{totem}.}
\begin{center}
\begin{tabular}{c|c|c|c}
\hline\hline
$\sqrt{s}$& A = 1/2 & A = 1 & TOTEM\\
\hline
7.0 TeV&  12.827$\pm$0.047 &  12.821$\pm$0.024 & 12.56$\pm$0.59\\
\hline
14 TeV & 13.811$\pm$0.068 & 13.903$\pm$0.033 & $-$\\
\hline\hline
\end{tabular} 
\end{center}
\end{table}

\section{Conclusions and Final Remarks}

Based on unitarity arguments and fits to the experimental data on the ratio
$R = \sigma_{el}/\sigma_{tot}$ from $pp$ scattering above 10 GeV, a novel empirical 
parametrization for $R(s)$ has been introduced, Eq. (17). The approximate
connection between this quantity and the ratio $\sigma_{tot}/B$, Eq. (16), allows us
to infer the corresponding energy dependence for this ratio in an almost model-independent way.
All the results are in agreement with rigorous inequalities derived
from the axiomatic formulation.

Although depending on the unknown asymptotic limit represented by the constant $A$,
the results here presented lead to, at least, four main conclusions:

\begin{enumerate}

\item  
If the black-disk represents a reliable physical limit \cite{bh}, its saturation is very
far from presently available energies: $\sqrt{s}$ $\gtrsim$ $10^9$ GeV, from
Figure 2;

\item 
If the Froissart-Lukaszuk-Martin bound is saturated then in this region
$B(s) \sim \ln^2s$;

\item  
Either for $A$ = 1/2 or $A$ = 1, the uncertainty bands evaluated by error
propagation from the fit parameters in Eq. (17) are negligible: upper, central and
lower curves in Figures 2, 3 and 4 overlap;

\item  
Even with $A$ = 1/2 and $A$ = 1 as lower and upper bounds,
extrapolation of the ratio $\sigma_{tot}/B$ to the Auger energy region,
$\sqrt{s} \sim$ 50 - 60 TeV, indicates a reasonably small error band,
overestimated from Figure 4 to be in the interval 15.5 $-$ 16.3.

\end{enumerate}

At last, we understand that the applicability of our results in the context
of the Glauber formalism can be further developed and improved along the following lines:

\begin{itemize}

\item
The recent TOTEM result for $\sigma_{el}$ at 7 TeV has been obtained through the
steps outlined in Subsection III.2 \cite{totem} and therefore has been evaluated from the
results for $\sigma_{tot}$ and $B$. The forthcoming measurement of $\sigma_{tot}$,
by means of a luminosity-independent method, and the corresponding $\sigma_{el}$
determination may improve our fit result, since this region is just associated
with the change of curvature in parametrization (17), as shown in Fig. 2. Moreover,
further results at 14 TeV will certainly contribute with additional improvements in the
fit results.

\item
Here we have limited the discussion to the extreme values indicated from unitarity and the
black-disk limit, together with 
the almost model-independent result for the ratio $\sigma_{tot}/B$.
However, as commented in Sect. III, analytical parametrizations for the  total cross
section and the $\rho$ parameter,
as those obtained by the COMPETE Collaboration (or possible deviation from this
result, if confirmed by the experimental data \cite{fms11}), may be combined with our results 
in the Glauber connection, Eqs. (8-9), reducing the uncertainties bands in the extrapolations
to cosmic-ray energies. 

\end{itemize}

\section*{Acknowledgments}

We are thankful to C. Dobrigkeit for discussions and a critical reading of the
manuscript.
Research supported by FAPESP 
(Contracts Nos. 11/00505-0, 09/50180-0).\\

\noindent
\textit{Note added}\\ 
After this paper was submitted for publication, we have noticed 
the results from a gray-disk-model analysis, in which a different 
transition on $\sigma_{el}/\sigma_{tot}(s)$, from low to high 
energies, is proposed \cite{ddd} (see also references therein).

\appendix
\section{}

Beyond the forward direction we can define

\begin{eqnarray}
B(s,t) =  \frac{d}{dt} \left[\ln \frac{d\sigma}{dt}(s,t) \right] 
\end{eqnarray}
and
\begin{eqnarray}
\rho\,(s,t) = \frac{{\rm Re} F(s, t)}{{\rm Im} F(s, t)},
\nonumber
\end{eqnarray}
so that from Eq. (1),
\begin{eqnarray}
B(s,t) =  2\frac{d}{dt} \ln \mathrm{Im} F(s,t) + \frac{d}{dt} \ln [1 + \rho^2(s,t)].
\end{eqnarray}
Under the reasonable assumption that at least in the \textit{neighborhood of} $t$=0 
\begin{eqnarray}
{\rm Im} F(s,\ t) \geq {\rm Re} F(s,t)
\nonumber
\end{eqnarray}
and by expanding the second term in the r.h.s of (A2) we obtain at $t$=0
\begin{eqnarray}
\left. \frac{d}{dt} \ln [1 + \rho^2(s,t)]  \right|_{t=0} =
\left.  2 \rho(s) \frac{d}{dt}\rho(s,t)  \right|_{t=0} + \mathrm{\cal{O}}(\rho^3(s)).
\nonumber
\end{eqnarray}
Since from the experimental data  $\rho(s) \lesssim$ 0.14 and under the assumption
\begin{eqnarray}
\lim_{t\rightarrow 0} \frac{d}{dt} \rho(s,t) = 0,
\nonumber
\end{eqnarray}
Eq. (A1) at $t$ = 0 reads
\begin{eqnarray}
B(s) \approx \left. 2 \frac{d}{dt} \ln \mathrm{Im} F(s,t) \right|_{t=0},
\nonumber
\end{eqnarray}
leading through Eq. (12) to the upper bound (13).

\end{document}